\documentclass[journal]{IEEEtran}

\newtheorem{dfntn}{Definition}

\usepackage{url}

\usepackage{graphicx}

\begin{document}

\title{The Key Authority -- Secure Key Management in Hierarchical Public Key Infrastructures}

\author{Alexander Wiesmaier, Marcus Lippert, Vangelis Karatsiolis\\
Technische Universit\"at Darmstadt\\
Department of Computer Science\\
Hochschulstr. 10; D-64289 Darmstadt, Germany\\
Tel: +49-6151-164889; Fax: +49-6151-166036}%

\maketitle
\thispagestyle{empty}
\pagestyle{empty}

\begin{abstract}
We model a private key's life cycle as a finite state machine. The
states are the key's phases of life and the transition functions
describe tasks to be done with the key. Based on this we define
and describe the key authority, a trust center module, which
potentiates the easy enforcement of secure management of private
keys in hierarchical public key infrastructures. This is done by
assembling all trust center tasks concerning the crucial handling
of private keys within one centralized module. As this module
resides under full control of the trust center's carrier it can
easily be protected by well-known organizational and technical
measures.
\end{abstract}


\section{Introduction}
This section introduces the reader into the matter of this paper. At
first, the scope of this paper is motivated. Then the problem we
address with this paper is stated. Finally, some related work is
presented.

\subsection{Motivation}
Nowadays a significant and growing contingent of communication is
done electronically over (often open) computer networks. In many
cases, it is necessary to guarantee the confidentiality, the
integrity and the authenticity of the thereby incurring electronic
data \cite{Tan02}. This can be achieved most reasonably by means
of hierarchical public key infrastructures \cite{HP02}.

\subsection{Problem}
Following \cite{AL99} the public key infrastructure itself must be
secure in order to provide a proper degree of security.
Consistently with \cite{DH76} we assume that the use of the
private key is necessary and sufficient for the decryption of a
cipher. Of course, we assume the same for the creation of a
digital signature. Thus, the handling of private keys is the most
crucial domain within a public key infrastructure. The problem
addressed in this paper is how to achieve secure management of
private keys within hierarchical public key infrastructures.

\subsection{Related work}
In the last years, there were many publications concerning topics
around public key infrastructures.  A lot of them deal with
concerns of certification \cite{KM99, Sar00, HL02} and revocation
\cite{NN98, GT00, AMLJK00}. An other often attended field is the
topic of trust and its chaining \cite{Lin00, BFL96, Mau96}. It was
recognized that the enforcement of revocations and policies is
achieved most easily in hierarchical public key infrastructures
\cite{Lin00}. However, there seems to be no publication dealing
with how to achieve secure key management within a hierarchical
public key infrastructure. Except a couple of proposed security
measures\footnote{e.g. those associated with the German "signature
act" \cite{regtp98techkomp, bsi98bsikat, regtp98zertst}}, there
appears to be no relevant work.

\section{Private Keys}
This section looks at the main topic of this paper: Private keys
within a hierarchical public key infrastructure. Firstly, the life
cycle of private keys is considered. Then we care about own and
foreign private keys. Lastly, we look at the security of private
keys.

\subsection{Life cycle}
\begin{figure}[htbp]
\centering
\includegraphics[width=\columnwidth]{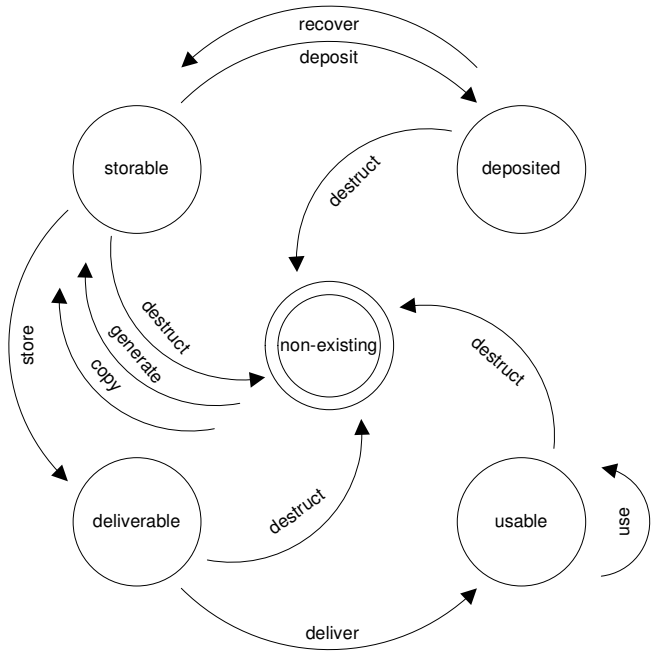}
\caption{A private key's life cycle\label{keyLifeCycle}}
\end{figure}

Figure \ref{keyLifeCycle} shows a simplified life cycle of a private
key within a public key infrastructure. Other thinkable life cycles
may contain more or less steps as well as other connections.
However, here the point of interest is not the exact life cycle but
an overview over the stations in a private key's life. Thus,
simplified here does not mean that some things were skipped. It
means that things are as concrete as necessary but as abstract as
possible in order to reflect a common life cycle in a preferably
simple diagram. For example, you can think of a special key
generation algorithm with possible private key recovery from the
public key. This can be viewed as being a conventional key
generation algorithm with mandatory key backup.

We model the life cycle of a private key as a finite state
machine. Each phase in a private key's life is represented by an
appropriate state of the machine. Having these states, it is easy
to construct the transition function. We interpret those
transitions as the tasks to be done with a private key within a
public key infrastructure. One instance of the finite state
machine belongs to exactly one instance of a private key. If we
have more than one instances of the same key each of them is
assigned its own finite state machine. The following paragraphs
explain the various states and transition functions in detail and
look at their security aspects.

\subsubsection{States}

\paragraph{Non-existing}
This is the initial and at the same time the final state of our
finite state machine. This is a virtual state. Being in this
state, means that the private key was destroyed or that it has
never existed at all. In this state, a private key is absolutely
secure, at the same time the key is also absolutely useless, since
it does not exist.

\paragraph{Storable}
The first real state in a private key's life is being storable.
This means the key was just generated, copied or recovered and
exists in the memory of a computer, a smartcard or any other
hardware. Now the key has to be stored or deposited somewhere to
make it persistent. This state is very dangerous, and the
unprotected key has to be shielded against eavesdropping or
manipulating.

\paragraph{Deposited}
For purposes of backup or similar cases, the private key can be
deposited. The key remains inactive in this state until it is
recovered. The deposited key may have to persist over a long time.
It must be easy for authorized parties to recover the private key,
while this must be impossible for unauthorized parties.

\paragraph{Deliverable}
To be delivered to the participant the private key should be
stored in a cryptographic token. Usually the token is personalized
and specially secured in this state. Being in this state the
private key should be properly protected by the token. As these
tokens are usually protected by secret pass phrases it is
necessary for them to remain a secret. In addition, the token must
be shielded against manipulating or interchanging.

\paragraph{Usable}
This is the really intended state of each private key. Now the key
is hold by its participant, who is able to use it. Here the key
also has to be protected against eavesdropping and manipulating.
This is particular difficult because the key must be handled in
clear for using it and the "unqualified" end user is involved.

\subsubsection{Transition functions}

\paragraph{Generate}
The first thing in a new private keys life cycle is always its
generation. For gaining good keys, the key parameters have to be
chosen carefully. Choosing bad parameters may result in gaining
insecure keys. In addition, a good random number generator has to
be used. Using a bad random generator may result in obtaining
predictable keys. Further, the key generation unit has to be
shielded against eavesdropping or manipulation. Thus, generating
keys is a very sensitive step and has to be done with expertise in
a suitable environment.

\paragraph{Copy}
Sometimes it is desirable to have more than one instance of the
same key. Thus, beside the generation of a fresh key we have the
possibility to construct a new instance by copying it from an
existing one. Copying private keys is somewhat delicate because
copying a private key is usually what is to be prevented.
Depending on the kind of utilized token it is not in all or even
in none state possible to make a copy of the private key. If the
copying unit handles unprotected private keys or pass phrases it
has to be shielded like the key generation unit.

\paragraph{Deposit}
Depositing a private key is not as easy as it seems to be at first
glance. The key must be deposited in a way that enables a long
period of persistence and at the same time features a maximum of
access protection. It is not suitable to just save it somewhere. A
strong access protection method is to be applied to the key and it
must be stored in a persistence-guaranteeing manner.

\paragraph{Recover}
Recovering a deposited key is the third and last method for
constructing an instance of a private key. Just like generating
and copying a key, the recovery has to be shielded against
eavesdropping or manipulation. It must be guaranteed that
authorized parties can easily recover the desired key, while this
must be impossible for unauthorized parties.

\paragraph{Store}
To make the key deliverable it should be stored in some kind of
cryptographic token, which is usually protected by secret pass
phrases. For storing purposes, it is necessary to handle the
private key and the pass phrases in clear text. Thus, appropriate
shielding of the storing step is demanded. For generating the pass
phrases the same demands concerning the choice of parameters and
random hardware apply as with the key generation.

\paragraph{Deliver}
As the private key should eventually reach the participant, it
must be delivered to him. Hereby it is necessary that the token
surely and solely reaches the intended participant. Further, it
must be ensured that no one is able to eavesdrop on the key or
manipulate or intercept the token. Delivering the respective pass
phrase must also be subject to massive shielding measures. See
also \cite{KLW04} for more work on this.

\paragraph{Use}
Using the private key is the really intended sense for having it.
However, it must be ensured that the authorized participant can
easily use the key while this must be impossible for all other
parties. Further, the key must be protected from eavesdropping,
manipulating or interchange.

\paragraph{Destruct}
The last thing in a private key's life is its destruction.
Although destructing a key seems to be easy there are some things
to be attended. If the key is to be destroyed, it is necessary to
do it in a way that the key cannot be restored anymore. Thus,
throwing away the token or simple file deletion from the hard disk
are no appropriate ways to destruct a private key.

\subsection{Own and Foreign Private Keys}
As we will have to distinguish between own and foreign private
keys later, these terms are defined now.

\begin{dfntn}\label{ownPrivKey}
A private key is called a participant's \emph{own private key} if
and only if the participant is entitled to use this private key.
\end{dfntn}

Usually the participant, which is associated with the
corresponding public key\footnote{e.g. in the certificate}, is
entitled to use the private key and thus, is its owner. However,
sometimes it can be a quiet different participant. If, for
example, an employee leaves the company, the superior might be the
new legal owner of the employee's decryption key to be able to
decrypt the employee's business documents.

\begin{dfntn}\label{foreignPrivKey}
A private key is a \emph{foreign private key} to a participant if
and only if it is not the participant's own private key.
\end{dfntn}

Thus, being the owner of a private key and being foreign to a
private key are mutually exclusive. At any given point in time and
for any given combination of a participant and a private key
exactly one of these predicates is true.

\subsection{Security of private keys}
Following \cite{DH76} the private key is sufficient and necessary
for decrypting a cipher or creating a signature. Thus, having
secure private keys is a necessary condition for having a secure
public key infrastructure. However, it is not a sufficient
condition for that. If the involved processes like registering the
participants or the communication with and within the trust center
are unreliable, the whole public key infrastructure is insecure.
However, those problems are not addressed here. Our scope is only
the secure management of private keys within a hierarchical public
key infrastructure.

The basic idea for securing private keys is that foreign private
keys are nobody's concern. However, as seen above there are some
stations and tasks concerning a private key, which require
attention, expertise and a suitable environment. We cannot expect
the ordinary public key infrastructure end user fulfilling these
requirements \cite{B}. Thus, it is better to give all those tasks
to one party and control it adequate. Following this approach, we
will define a trust center module called key authority (KA), which
potentiates the easy enforcement of secure management of private
keys.

Depending on the used technology and policy, more or less tasks
have to be fulfilled by KA. When using high-grade smart cards with
key generation features and having no key backup the KA is
concerned only with the issuer private keys for signing the
certificate. When using soft tokens and key backup the KA is
concerned in addition with the full range of user key generation,
deposition, recovery, and so on. Even if the KA does not execute
certain tasks, it can at least increase their security by
prearranging those tasks. An example for this case is the usage of
the private key. This is naturally a task to be done at the end
user's side. However, KA can enforce its security by issuing
active tokens, which do not reveal the private key at all.

\section{Core Key Authority}
In order to comprehend the specification of the key authority it
is necessary to have a common vocabulary. Thus, this section gives
definitions of some relevant terms, which are new or might be
mistakable. Further, it introduces some simplifications to ease
the definition of the key authority.

\subsection{Issuer}
In hierarchical public key infrastructures, there is at least one
dedicated participant, who issues the certificates. Each of those
participants has at least one distinguished name and at least one
private key for issuing the certificates. Further, each of those
participants may respectively issue certificates using several
distinguished names and private keys. In order to have clear
circumstances we define the term issuer as follows.

\begin{dfntn}\label{issuer}
An \emph{issuer} is a participant within a hierarchical public key
infrastructure, which has exactly and exclusively one
distinguished name. This participant has exactly and exclusively
one valid private key for issuing certificates at a point of time.
\end{dfntn}

Thus, each of the mentioned three components (issuer,
distinguished name and valid private key) does non-ambiguously
determine the respective two others. An issuer may have more tasks
than issuing certificates\footnote{e.g. revoking certificates} and
several private keys for accomplishing those tasks. Further, an
issuer might use several private keys for certifying over the
time, but only one of them will be valid at a point of time.

\subsection{Core Public Key Infrastructure}
A public key infrastructure may consist of a number of smaller sub
public key infrastructures, which are connected by means of
bridging, cross certifying or somehow else. Those sub public key
infrastructures in turn may consist of smaller sub public key
infrastructures and so on. This can lead to complicated
relationships between the involved trust centers, issuers,
participants and their keys. In order to avoid misunderstandings
in the further, discussion we introduce to the following
simplification of a public key infrastructure.

\begin{dfntn}\label{spannedPki}
An issuer's \emph{core public key infrastructure} is part of a
whole public key infrastructure. It contains exactly those trust
center products\footnote{key pairs, certificates, tokens, etc.},
which were issued by the respective issuer. Further, it consists
of exactly those participants and clients, which deal with those
trust center products.
\end{dfntn}

Thus, every issuer has exactly one core public key infrastructure
and every core public key infrastructure belongs to exactly one
issuer.

\subsection{Defining the Core Key Authority}
We are now ready to define the core key authority. This is done by
defining, which tasks the authority has within the core public key
infrastructure.

\begin{dfntn}
An issuer's \emph{core key authority} executes exactly and
exclusively those tasks within the core public key infrastructure
which require or enable the access to the issuer's own private
keys or to any foreign private keys.
\end{dfntn}

All other parties are not allowed to get in touch with any private
keys except their own ones. Although the key authority is in
principle allowed to handle foreign private keys, the regarding
owners of those might\footnote{according to the trust center's
policy} deny the access to their private keys. Thus, the motto is
"If someone is allowed to see foreign private keys, it is the core
key authority and no one else".

\subsection{Tasks}
The above definition of the core key authority gives a general
description of its tasks. This subsection gives a more concrete
specification of them. We describe all tasks, which require or
enable the access to the issuer's or foreign private keys.

\subsubsection{Issuing}
The core key authority issues certificates on behalf of its
issuer. Issuing the certificates means signing them with the
issuer's appropriate private key.

\subsubsection{Revocation}
If it is the issuer's task to revoke certificates this is done by
its core key authority, too. Revoking a certificate means signing
a certificate revocation list\footnote{or a similar data
structure} containing that certificate with the appropriate
private key.

\subsubsection{Remaining Private Keys}
The issuer might have additional private keys for accomplishing
various tasks. All those tasks will be done by the issuer's core
key authority because no one else is allowed to use the issuer's
private keys.

\subsubsection{Key Generation}
The core key authority generates all key pairs, which are not
generated by their respective owners for themselves. This is
because generating keys implies the access to them, which is
granted only to the owner and the key authority. Of course, the
key authority generates the respective issuer's key pairs.

\subsubsection{Personalization}
If the core key authority has generated a key pair, it has to
store the private key within a personal security environment. As
personal security environments are protected by (strong) pass
phrases\footnote{or similar mechanisms} it is also the core key
authority's task to generate them and to inform the regarding
owner about them. This is the key authority's task as possessing
the personal security environment and knowing the pass phrase
enables the access to the stored keys.

\subsubsection{Archiving/Backup/Recovery}
When a private key has to be archived, backuped or recovered it is
the task of the core key authority to do so if the respective
owner does not do this. Any other parties are not allowed to do
this, as they are not allowed to see foreign private keys.

\subsection{Secure Key Management}
This sections shows why the core key authority defined above
enables an easy enforcement of secure private key management
within a core public key infrastructure by enabling the leverage
of suitable technical and organizational measures.

\subsubsection{Easy Enforcement of Security}
As the core key authority is the only party within the core public
key infrastructure, which has access to the issuer's private key
or to foreign private keys, one can protect those keys by just
protecting the core key authority. This is manageable because you
have to protect only one party instead of many. Further, it is
uncomplicated because this party is located within a known
environment\footnote{at the trust center} instead of somewhere in
any participant's environment. Thus, it is easy to enforce secure
key management with suitable (and well-known) technical and
organizational measures.

\subsubsection{Technical Measures}
As all crucial private keys are under full control of the carrier
of the core public key infrastructure\footnote{by being assembled
within the core key authority}, it is possible to use all kinds of
common technical measures for protecting keys. Examples for those
measures are physical shielding, cryptographic hardware and more.

\subsubsection{Organizational Measures}
As with the technical measures, the use of common organizational
measures for protecting the private keys is enabled by the
centralized maintenance of crucial tasks. Those measures include
running in offline mode, dual control\footnote{always two
operators must be logged in} and so on.

\section{Key Authority}
This section explains how an arbitrary public key infrastructure
can be seen as a composite of several core public key
infrastructures. Further, it is shown that the security achieved
by using a core key authority\footnote{within a core public key
infrastructure} is automatically passed to a composite public key
infrastructure.

\subsection{Public Key Infrastructures}
\begin{figure}[htbp]
\centering
\includegraphics[width=\columnwidth]{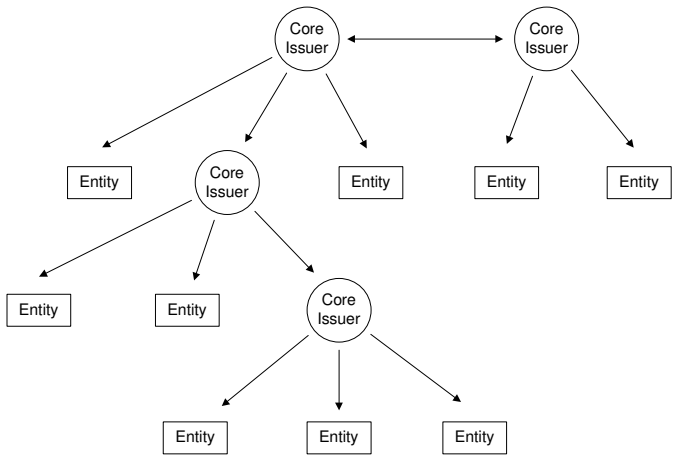}
\caption{A PKI as composition of core PKIs\label{composedPki}}
\end{figure}

Any (conventional) hierarchical public key infrastructure can be
seen as a collection of connected core public key infrastructures.
From the view of the core public key infrastructure, the "other"
issuers appear as normal participants. From the view of the whole
public key infrastructure, the various core issuers appear as
cross-certified or bridged sub public key infrastructures. Figure
\ref{composedPki} shows an example of a composed public key
infrastructure tree.

\subsection{Key Authority}
The various core key authorities may be operated within one
module. This module is exclusively for assembling several core key
authorities and is called \emph{key authority}. It provides the
environment to share the technical and organizational measures
between the miscellaneous core key authorities. Each core key
authority is operated within exactly one key authority and each
key authority contains at least one core key authority. If there
is exactly one core key authority within a key authority the two
terms are synonymous.

\subsection{Security}
All tasks concerning issuer private keys or foreign private keys
are conducted within a key authority. As shown above it is easy to
protect (core) key authorities. Thus, we can easily enforce the
secure key management within the whole public key infrastructure.

\section{Conclusion}
This paper introduced a finite state machine model for a private
keys life cycle. This model was used to identify the various tasks
concerning private keys. Based on this the key authority, a trust
center module, which performs all tasks concerning private issuer
keys or foreign private keys was introduced. It was shown that
using this module allows for an easy enforcement of secure key
management within a hierarchical public key infrastructure.

\bibliographystyle{IEEEtran}
\bibliography{myReferences}

\end{document}